\documentclass[twocolumn,showpacs,preprintnumbers,pra,hyperref]{revtex4}
\usepackage{amssymb}
\usepackage{amsfonts}
\usepackage{amsmath}
\usepackage{graphicx}

\begin{document}

\title{The effect of defect layer on transmissivity and light field distribution in general function photonic crystals}
\author{Xiang-Yao Wu$^{a}$\thanks{%
E-mail: wuxy2066@163.com}, Si-Qi Zhang$^{a}$, Bo-Jun Zhang$^{a}$, Xiao-Jing Liu$^{a}$, \\
 Jing Wang$^{a}$, Hong Li$^{a}$, Nuo Ba$^{a}$ and Xin-Guo Yin$^{b}$} \affiliation{$^{a}${\small
Institute of Physics, Jilin Normal University, Siping 136000, China}\\
$^{b}${\small  Institute of Physics, Xuzhou Normal University,
Xuzhou 221000, China} }

\begin{abstract}
We have theoretically investigated a general function photonic
crystals (GFPCs) with defect layer, and choose the line refractive
index function for two mediums $A$ and $B$, and analyze the effect
of defect layer's position, refractive indexes and period numbers
on the transmission intensity and the electric field distribution.
We obtain some new characters that are different from the
conventional PCs, which should be helpful in the design of
photonic crystals.

\end{abstract}

\maketitle

\maketitle {\bf 1. Introduction} \vskip 8pt

Photonic crystals (PC) are a new kind of materials which
facilitate the control of the light [1]. PC exhibits Photonic Band
Gaps (PBG) that forbids the radiation propagation in a specific
range of frequencies [2-6]. The PBG forbids the radiation
propagation in a specific range of frequencies. The existence of
PBGs will lead to many interesting phenomena, e.g., modification
of spontaneous emission [7-9] and photon localization [10]. Thus
numerous applications of photonic crystals have been proposed in
improving the performance of optoelectronic and microwave devices
such as high-efficiency semiconductor lasers, right emitting
diodes, wave guides, optical filters, high-Q resonators, antennas,
frequency-selective surface, optical limiters and amplifiers
[11-18]. In the past ten years has been developed an intensive
effort to study and micro-fabricate PBG materials in one, two or
three dimensions [19-21]. In Refs. [22-25], we have proposed a
general function photonic crystals (GFPCs), which refractive index
is a arbitrary function of space position. Unlike conventional
photonic crystals (PCs), which structure grow from two materials,
A and B, with different dielectric constants $\varepsilon_{A}$ and
$\varepsilon_{B}$, and have obtained some results different from
the conventional photonic crystals. In the paper, We have studied
the general function photonic crystals (GFPCs) with defect layer,
and choose the line refractive index function for two mediums $A$
and $B$. We obtain some results: (1) When the position of defect
layer move behind in the GFPCs, or the refractive indexes of
defect layer increase, the transmission intensity maximum of
defect model decreases. (2) When the period number of GFPCs
increase, the transmission intensity of defect model increase, and
the defect model's width of half height become narrow. (3) Both
the defect layer and it's position have the effects on the
electric field. (4) In the structure $(BA)^{N}D(BA)^{N}$, when the
period number $N$ increase, the relative intensity of electric
field increase. (5) When the refractive indexes of defect layer
($n_{d}$) increase, the relative intensity of electric field was
enhanced.

\vskip 8pt

{\bf 2. The light motion equation in general function photonic
crystals} \vskip 8pt

For the general function photonic crystals, the medium refractive
index is a periodic function of the space position, which can be
written as $n(z)$, $n(x, z)$ and $n(x, y, z)$ corresponding to
one-dimensional, two-dimensional and three-dimensional function
photonic crystals. In the following, we shall deduce the light
motion equations of the one-dimensional general function photonic
crystals, i.e., the refractive index function is $n=n(z)$,
meanwhile motion path is on $xz$ plane. The incident light wave
strikes plane interface point $A$, the curves $AB$ and $BC$ are
the path of incident and reflected light respectively, and they
are shown in FIG. 1.
\begin{figure}[tbp]
\includegraphics[width=8.5 cm]{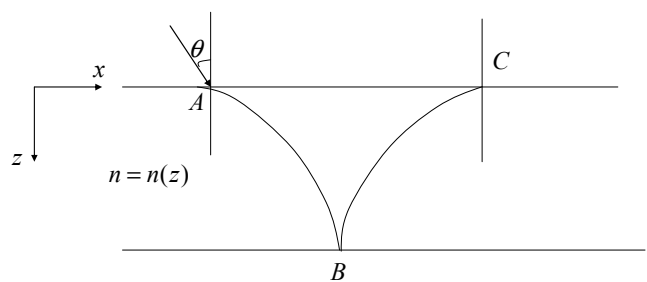}
\caption{The motion path of light in the medium of refractive
index $n(z)$.}
\end{figure}

The light motion equation can be obtained by Fermat principle, it
is
\begin{eqnarray}
\delta\int^{B}_{A}n(z) ds=0.\label{1}
\end{eqnarray}
In the two-dimensional transmission space, the line element $ds$
is
\begin{eqnarray}
ds=\sqrt{(dx)^{2}+(dz)^{2}}=\sqrt{1+\dot{z}^{2}}dx,
\end{eqnarray}
where $\dot{z}=\frac{dz}{dx}$, then Eq. (1) becomes
\begin{eqnarray}
\delta\int^{B}_{A}n(z)\sqrt{1+(\dot{z})^{2}}dx=0.
\end{eqnarray}
The Eq. (3) change into
\begin{eqnarray}
\int^{B}_{A}(\frac{\partial(n(z)\sqrt{1+\dot{z}^{2}})}{\partial
z}\delta z+\frac{\partial(n(z)\sqrt{1+\dot{z}^{2}})}{\partial
\dot{z}}\delta\dot{z})dx=0,
\end{eqnarray}
At the two end points $A$ and $B$, their variation is zero, i.e.,
$\delta z (A)=\delta z (B)=0$. For arbitrary variation $\delta z$, the Eq. (4) becomes \\
\begin{eqnarray}
&&\frac{dn(z)}{dz}\sqrt{1+\dot{z}^{2}} -\frac{d n(z)}{d
z}\dot{z}^{2}(1+\dot{z}^{2})^{-\frac{1}{2}}
\nonumber\\&&-n(z)\frac{\ddot{z}\sqrt{1+\dot{z}^{2}}
-\dot{z}^{2}\ddot{z}(1+\dot{z}^{2})^{-\frac{1}{2}}}{1+\dot{z}^{2}}
=0,
\end{eqnarray}
simplify Eq. (5), we have
\begin{eqnarray}
\frac{d n(z)}{n(z)} = \frac{\dot{z}d\dot{z}}{1+\dot{z}^{2}}.
\end{eqnarray}\\
The Eq. (6) is light motion equation in one-dimensional function
photonic crystals.
\begin{figure}[tbp]
\includegraphics[width=8.5 cm]{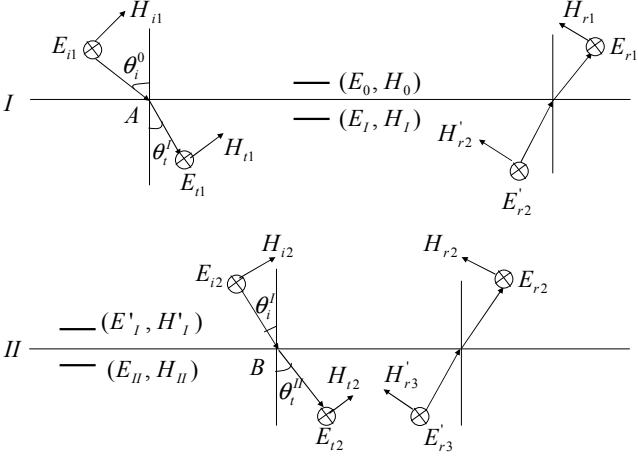}
\caption{The light transmission and electric magnetic field
distribution figure in FIG.1 medium.} \label{Fig1}
\end{figure}
\vskip 8pt

{\bf 3. The transfer matrix of one-dimensional general function
photonic crystals} \vskip 8pt

In this section, we should calculate the transfer matrix of
one-dimensional general function photonic crystals. In fact, there
is the reflection and refraction of light at a plane surface of
two media with different dielectric properties. The dynamic
properties of the electric field and magnetic field are contained
in the boundary conditions: normal components of $D$ and $B$ are
continuous; tangential components of $E$ and $H$ are continuous.
We consider the electric field perpendicular to the plane of
incidence, and the coordinate system and symbols as shown in FIG.
2.

On the two sides of interface I, the tangential components of
electric field $E$ and magnetic field $H$ are continuous, there
are

\begin{eqnarray}
\left \{ \begin{array}{ll}
 E_{0}=E_{I}=E_{t1}+E'_{r2}\\
H_{0}=H_{I}=H_{t1}\cos\theta_{t}^{I}-H'_{r2}\cos\theta_{t}^{I}.
\end{array}
\right.
\end{eqnarray}
On the two sides of interface II, the tangential components of
electric field $E$ and magnetic field $H$ are continuous, and give
\begin{eqnarray}
\left \{ \begin{array}{ll}
 E_{II}=E'_{I}=E_{i2}+E_{r2}\\
H_{II}=H'_{I}=H_{i2}\cos\theta_{i}^{I}-H_{r2}\cos\theta_{i}^{I},
\end{array}
\right.
\end{eqnarray}
the electric field ${E_{t1}}$ is
\begin{eqnarray}
E_{t1}=E_{t10}{e^{i(k_{x}x_{A}+k_{z}z)}|_{z=0}}=E_{t10}e^{i\frac{\omega}{c}n(0)\sin\theta_{t}^{I}x_{A}},
\end{eqnarray}
and the electric field ${E_{i2}}$ is
\begin{eqnarray}
E_{i2}&=&E_{t10}{e^{i(k'_{x}x_{B}+k'_{z} z)}|_{z=b}}
\nonumber\\&=&E_{t10}e^{i\frac{\omega}{c}n(b)(\sin\theta_{i}^{I}x_{B}+\cos\theta_{i}^{I}
b)}.
\end{eqnarray}
Where $x_{A}$ and $x_{B}$ are $x$ component coordinates
corresponding to point $A$ and point $B$. We should give the
relation between $E_{i2}$ and $E_{t1}$. By integrating the two
sides of Eq. (6), we can obtain the coordinate component $x_{B}$
of point $B$
\begin{eqnarray}
\int^{n(z)}_{n(0)}\frac{dn(z)}{n(z)}=\int^{k_{z}}_{k_{0}}\frac{\dot{z}d\dot{z}}{1+\dot{z}^{2}},
\end{eqnarray}
to get
\begin{eqnarray}
k_z^2=(1+k_0^2)(\frac{n(z)}{n(0)})^2-1,
\end{eqnarray}
and
\begin{eqnarray}
dx=\frac{dz}{\sqrt{(1+k_{0}^{2})(\frac{n(z)}{n(0)})^{2}-1}}.
\end{eqnarray}
where $k_{0}=\cot\theta_{t}^{I}$ and $k_{z}=\frac{d z}{d x}$ From
Eq. (12), there is $n(z)>n(0)\sin\theta^{I}_{t}$. and the
coordinate $x_{B}$ is
\begin{eqnarray}
x_{B}=x_{A}+\int^{b}_{0}\frac{dz}{\sqrt{(1+k_{0}^{2})(\frac{n(z)}{n(0)})^{2}-1}},
\end{eqnarray}
where $b$ is the medium thickness of FIG. 1 and FIG. 2.\\
By substituting Eqs. (9) and (14)into (10), and using the equality
\begin{eqnarray}
 n(0)\sin\theta_{t}^{I}=n(b)\sin\theta_{i}^{I},
\end{eqnarray}
  we have
\begin{eqnarray}
E_{i2}&=&E_{t1}e^{i{\delta}_{b}},
\end{eqnarray}
where
\begin{figure}[tbp]
\includegraphics[width=8 cm]{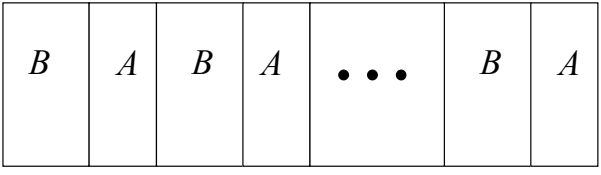}
\caption{The structure $(BA)^{N}$ of the general function photonic
crystals.}
\end{figure}
\begin{eqnarray}
\delta_{b}=\frac{\omega}{c}n_{b}(b)(\cos\theta_{i}^{I}b+\sin\theta_{i}^{I}
\int^{b}_{0}\frac{dz}{\sqrt{\frac{n_{b}^{2}(z)}{n_{0}^{2}\sin^{2}\theta_{i}^{0}}-1}}),
\end{eqnarray}
and similarly
\begin{eqnarray}
E'_{r2}=E_{r2}e^{i\delta_{b}}.
\end{eqnarray}
Substituting Eqs. (16) and (18) into (7) and (8), and using
$H=\sqrt{\frac{\varepsilon_{0}}{\mu_{0}}}nE$, we obtain
\begin{eqnarray}
\left(%
\begin{array}{c}
  E_{I} \\
  H_{I} \\
\end{array}%
\right)&=&M_{B}\left(%
\begin{array}{c}
  E_{II} \\
  H_{II} \\
\end{array}%
\right),
\end{eqnarray}
where
\begin{eqnarray}
M_{B}=\left(%
\begin{array}{cc}
 \cos\delta_{b} & -\frac{i\sin\delta_{b}}{\sqrt{\frac{\varepsilon_{0}}{\mu_{0}}}n_{b}(b)\cos\theta_{i}^{I}} \\
 -in_{b}(0)\sqrt{\frac{\varepsilon_{0}}{\mu_{o}}}\cos\theta_{t}^{I}\sin\delta_{b}
 & \frac{n_{b}(0)\cos\theta_{t}^{I}\cos\delta_{b}}{n_{b}(b)\cos\theta_{i}^{I}}\\
\end{array}%
\right),
\end{eqnarray}
The Eq. (20) is the transfer matrix $M$ in the medium of FIG. 1
and FIG. 2. By refraction law, we can obtain
\begin{eqnarray}
\sin\theta^{I}_{t}=\frac{n_{0}}{n(0)}\sin\theta^{0}_{i},\cos\theta^{I}_{t}
=\sqrt{1-\frac{n_{0}^{2}}{n^{2}(0)}\sin^{2}\theta^{I}_{t}},
\end{eqnarray}
where $n_0$ is air refractive index, and $n(0)=n(z)|_{z=0}$. Using
Eqs. (15) and (21), we can calculate $\cos\theta_{i}^{I}$.

\begin{figure}[tbp]
\includegraphics[width=8 cm]{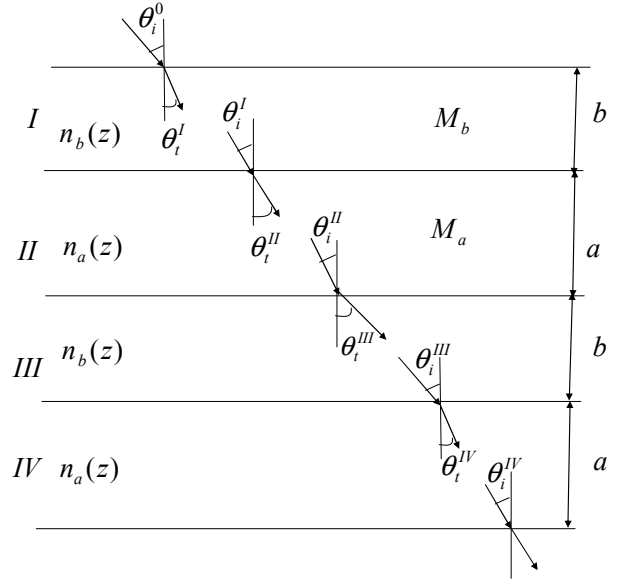}
\caption{The two periods transmission figure of light in general
function photonic crystals.}
\end{figure}
\vskip 8pt

{\bf 4. The transmissivity and light field distribution of
one-dimensional general function photonic crystals} \vskip 8pt

In section 3, we obtain the $M$ matrix of the half period. We know
that the conventional photonic crystals is constituted by two
different refractive index medium, and the refractive indexes are
not continuous on the interface of the two mediums. We could
devise the one-dimensional general function photonic crystals
structure as follows: in the first half period, the refractive
index distributing function of medium $B$ is $n_{b}(z)$. and in
the second half period, the refractive index distributing function
of medium $A$ is $n_{a}(z)$, corresponding thicknesses are $b$ and
$a$, respectively. Their refractive indexes satisfy condition
$n_{b}(b)\neq n_{a}(0)$, their structure are shown in FIG. 3, and
FIG. 4. The Eq. (20) is the half period transfer matrix of medium
$B$. Obviously, the half period transfer matrix of medium A is
\begin{eqnarray}
M_{A}=\left(%
\begin{array}{cc}
 \cos\delta_{a} & -\frac{i\sin\delta_{a}}{\sqrt{\frac{\varepsilon_{0}}{\mu_{0}}}n_{a}(a)\cos\theta_{i}^{II}} \\
 -in_{a}(0)\sqrt{\frac{\varepsilon_{0}}{\mu_{o}}}\cos\theta_{t}^{II}\sin\delta_{a}
 & \frac{n_{a}(0)\cos\theta_{t}^{II}\cos\delta_{a}}{n_{a}(a)\cos\theta_{i}^{II}}\\
\end{array}%
\right),
\end{eqnarray}
where
\begin{eqnarray}
\delta_{a}&=&\frac{\omega}{c}n_{a}(a)[\cos\theta^{II}_{i}\cdot a
\nonumber\\&&
+\sin\theta^{II}_{i}\int^{a}_{0}\frac{dz}{\sqrt{\frac{n_{a}^{2}(z)}{n_{0}^{2}\sin^{2}\theta_{i}^{0}}-1}}],
\end{eqnarray}
\begin{eqnarray}
\cos\theta^{II}_{t}
=\sqrt{1-\frac{n_{0}^{2}}{n_{a}^{2}(0)}\sin^{2}\theta_{i}^{0}},
\end{eqnarray}
and
\begin{eqnarray}
\sin\theta^{II}_{i}=\frac{n_{0}}{n_{a}(a)}\sin\theta_{i}^{0},
\end{eqnarray}
\begin{eqnarray}
\cos\theta^{II}_{i}
=\sqrt{1-\frac{n_{0}^{2}}{n_{a}^{2}(a)}\sin^{2}\theta_{i}^{0}}.
\end{eqnarray}
In one period, the transfer matrix $M$ is
\begin{eqnarray}
&&M=M_{B}\cdot M_{A}\nonumber\\
&&=\left(%
\begin{array}{cc}
  \cos\delta_{b} & \frac{-i\sin\delta_{b}}{\sqrt{\frac{\varepsilon_{0}}{\mu_{0}}}n_{b}(b)\cos\theta_{i}^{I}} \\
 -in_{b}(0)\sqrt{\frac{\varepsilon_{0}}{\mu_{o}}}\cos\theta_{t}^{I}\sin\delta_{b}
 & \frac{n_{b}(0)\cos\theta_{t}^{I}\cos\delta_{b}}{n_{b}(b)\cos\theta_{i}^{I}}\\
\end{array}%
\right) \nonumber\\&&
\left(%
\begin{array}{cc}
   \cos\delta_{a} & \frac{-i\sin\delta_{a}}{\sqrt{\frac{\varepsilon_{0}}{\mu_{0}}}n_{a}(a)\cos\theta_{i}^{II}} \\
 -in_{a}(0)\sqrt{\frac{\varepsilon_{0}}{\mu_{o}}}\cos\theta_{t}^{II}\sin\delta_{a}
 & \frac{n_{a}(0)\cos\theta_{t}^{II}\cos\delta_{a}}{n_{a}(a)\cos\theta_{i}^{II}}\\
\end{array}%
\right).
\end{eqnarray}
The defect layer's refractive index is constant $n_{d}$, its
transfer matrix $M_{d}$ is
\begin{eqnarray}
M_{d}=\left(%
\begin{array}{cc}
 \cos\delta_{d} & -\frac{i}{\eta_{d}}\sin\delta_{d} \\
 -i\eta_{d}\sin\delta_{d}
 & \cos\delta_{d}\\
\end{array}%
\right),
\end{eqnarray}
where $\eta_{d}=\sqrt{\frac{\varepsilon_{0}}{\mu_{0}}}n_{d}$, $
\delta_{d}=\frac{\omega}{c}n_{d}d$.\\
The form of the GFPCs transfer matrix $M$ is more complex than the
conventional PCs. The angle $\theta_{t}^{I}$, $\theta_{i}^{I}$,
$\theta_{t}^{II}$ and $\theta_{i}^{II}$ are shown in Fig. 4. The
characteristic equation of GFPCs is
\begin{eqnarray}
\left(%
\begin{array}{c}
  E_{1} \\
  H_{1} \\
\end{array}%
\right)&=&M_{1}M_{2}\cdot\cdot\cdot M_{d}\cdot\cdot\cdot M_{N}
\left(%
\begin{array}{c}
  E_{N+1} \\
  H_{N+1} \\
\end{array}%
\right) \nonumber\\&=&M_{b}M_{a}M_{b}M_{a}\cdot\cdot\cdot M_{d}\cdot\cdot\cdot M_{b}M_{a}\left(%
\begin{array}{c}
  E_{N+1} \\
  H_{N+1} \\
\end{array}%
\right)
\nonumber\\&=&M\left(%
\begin{array}{c}
  E_{N+1} \\
  H_{N+1} \\
\end{array}%
\right)=\left(%
\begin{array}{c c}
  A &  B \\
 C &  D \\
\end{array}%
\right)
 \left(%
\begin{array}{c}
  E_{N+1} \\
  H_{N+1} \\
\end{array}%
\right).
\end{eqnarray}
Where $N$ is the period number.
\begin{figure}[tbp]
\includegraphics[width=9 cm]{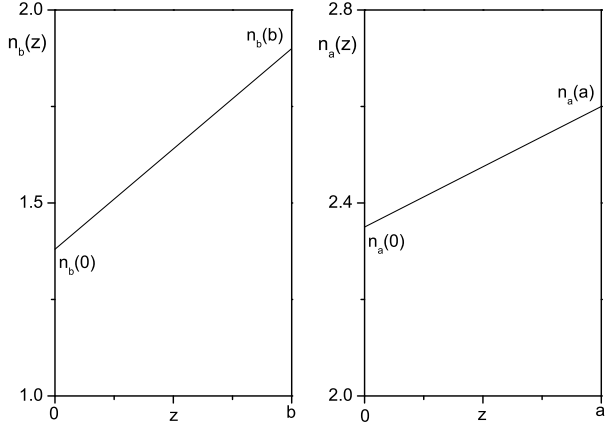}
\caption{The refractive index of the general functions in a
period.}
\end{figure}
\begin{figure}[tbp]
\includegraphics[width=9 cm]{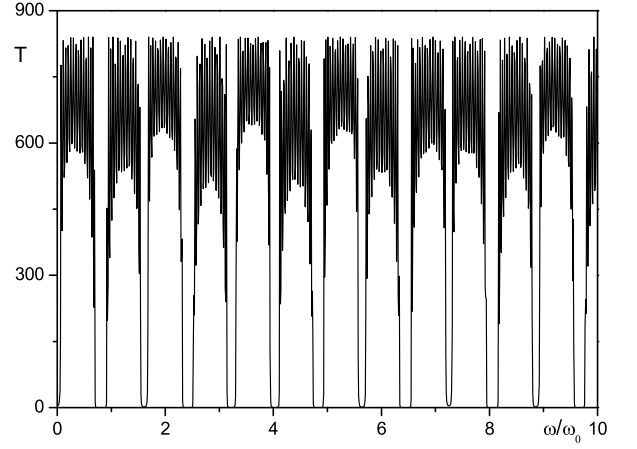}
\caption{The relation between transmissivity and frequency
corresponding to the general function photonic crystals.}
\end{figure}
With the transfer matrix $M$ (Eq. (28)), we can obtain the
transmission and reflection coefficient $t$ and $t$, and the
transmissivity and reflectivity $T$ and $R$, they are
\begin{eqnarray}
t=\frac{E_{tN+1}}{E_{i1}}=\frac{2\eta_{0}}{A\eta_{0}+B\eta_{0}\eta_{N+1}+C+D\eta_{N+1}},
\end{eqnarray}
\begin{eqnarray}
r=\frac{E_{r1}}{E_{i1}}=\frac{A\eta_{0}+B\eta_{0}\eta_{N+1}-C-D\eta_{0}}{A\eta_{0}+B\eta_{0}\eta_{N+1}+C+D\eta_{0}},
\end{eqnarray}
and
\begin{eqnarray}
T=t\cdot t^{*},
\end{eqnarray}
\begin{eqnarray}
R=r\cdot r^{*}.
\end{eqnarray}
Where $\eta_{0}=\eta_{N+1}=\sqrt{\frac{\varepsilon_0}{\mu_0}}$. In
the following, we give the electric field distribution of light in
the one-dimensional GFPCs. The propagation figure of light in
one-dimensional GFPCs is shown in FIG. 9. From Eq. (28), we have
\begin{eqnarray}
&&\left(%
\begin{array}{c}
  E_{1} \\
  H_{1} \\
\end{array}%
\right)=M_{1}(d_{1})M_{2}(d_{2})\cdot\cdot\cdot
M_{d}(d)\cdot\cdot\cdot M_{N-1}(d_{N-1}) \nonumber\\&&
M_{N}(\Delta z_{N})
\left(%
\begin{array}{c}
  E_{N}(d_{1}+d_{2}{\cdots}+d_{N-1}+\Delta z_{N}) \\
  H_{N}(d_{1}+d_{2}{\cdots}+d_{N-1}+\Delta z_{N}) \\
\end{array}%
\right)
\end{eqnarray}
where $d_1$ and $d_2$ are the thickness of first and second
period, respectively, $\Delta z_{N}$ is the propagation distance
of light in the N-th period, $E_1$ and $H_1$ are the intensity of
incident electric field and magnetic field, and
$E_{N}(d_{1}+d_{2}{\cdots}+d_{N-1}+\Delta z_{N})$ and
$H_{N}(d_{1}+d_{2}{\cdots}+d_{N-1}+\Delta z_{N})$ are the
intensity of the N-th period electric field and magnetic field.
The Eq. (34) can be written as
\begin{eqnarray}
&&\left(%
\begin{array}{c}
  E_{N}(d_{1}+d_{2}{\cdots}+d_{N-1}+\Delta z_{N}) \\
  H_{N}(d_{1}+d_{2}{\cdots}+d_{N-1}+\Delta z_{N}) \\
\end{array}%
\right) =M^{-1}_{N}(\Delta z_{N})\nonumber\\&&
M^{-1}_{N-1}(d_{N-1}) \cdot\cdot\cdot+ M^{-1}(d) \cdot\cdot\cdot
M^{-1}_{2}(d_{2}) M^{-1}_{1}(d_{1})
\left(%
\begin{array}{c}
  E_{1} \\
  H_{1} \\
\end{array}%
\right)\nonumber\\&&\left(%
\begin{array}{cc}
  A(\Delta z_{N}) & B(\Delta z_{N}) \\
  C(\Delta z_{N}) & D(\Delta z_{N}) \\
\end{array}%
\right)\left(%
\begin{array}{c}
  E_{1} \\
  H_{1} \\
\end{array}%
\right),
\end{eqnarray}
the electric field $E_1$ and magnetic field $H_1$ can be written
as
\begin{eqnarray}
E_{1}=E_{i1}+E_{r1}=(1+r)E_{i1},
\end{eqnarray}
\begin{eqnarray}
H_{1}&=&H_{i1}\cos\theta_{i}^{0}-H_{r1}\cos\theta_{i}^{0}
\nonumber\\&=&\sqrt{\frac{\varepsilon_{0}}{\mu_{0}}}\cos\theta_{i}^{0}(1-r)E_{i1}.
\end{eqnarray}
From Eqs. (34)-(36), we can obtain the ratio of the electric field
$E_{N}(d_{1}+d_{2}{\cdots}+d_{N-1}+\Delta z_{N})$ within the GFPCs
to the incident electric field $E_{i1}$, it is
\begin{eqnarray}
&&|\frac{ E_{N}(d_{1}+d_{2}{\cdots}+d_{N-1}+\Delta
z_{N})}{E_{i1}}|^{2}\nonumber\\&&=|A(\Delta z_{N})(1+r)+B(\Delta
z_{N})\sqrt{\frac{\epsilon_{0}}{\mu_{0}}}\cos\theta_{i}^{0}(1-r)|^{2}.
\end{eqnarray}
\vskip 2pt

{\bf 5. Numerical result}

\vskip 2pt

In this section, we report  our numerical results of
transmissivity and light field distribution. We consider
refractive indexes of the linearity functions in a period, it is

\begin{eqnarray}
n_{b}(z)=n_{b}(0)+\frac{n_{b}(b)-n_{b}(0)}{b}z, \hspace{0.1in} 0
\leq z\leq b,
\end{eqnarray}
\begin{eqnarray}
n_{a}(z)=n_{a}(0)+\frac{n_{a}(a)-n_{a}(0)}{a}z, \hspace{0.1in} 0
\leq z\leq a,
\end{eqnarray}
Eqs. (38) and (39) are the line refractive indexes distribution
functions of two half period mediums $B$ and $A$. When the
endpoint values $n_{b}(0)$, $n_{b}(b)$, $n_{a}(0)$ and $n_{a}(a)$
are all given, the line refractive index functions $n_{b}(z)$ and
$n_{a}(z)$ are ascertained. The main parameters are: the half
period thickness $b$ and $a$, the starting point refractive
indexes $n_{b}(0)$ and $n_{a}(0)$, and end point refractive
indexes $n_{b}(b)$ and $n_{a}(a)$, the optical thickness of the
two mediums are equal, i.e., $n_{b}(0)b=n_{a}(0)a$, the incident
angle $\theta_{i}^{0}=0$, the center frequency
$\omega_{0}=1.215\times10^{15} Hz$, the center wave length
$\lambda_{0}=\frac{2\pi c}{\omega_{0}}$, the thickness $b=280 nm$,
$a=165 nm$ and the period number $N=16$. we take $n_{b}(0)=1.38$,
$n_{b}(b)=1.9$ for the medium $B$, and $n_{a}(0)=2.35$,
$n_{a}(a)=2.6$ for the medium $A$, which are the up line function
of refractive indexes, it is shown in FIG.5. By the refractive
indexes function, we can calculate the transmissivity, we obtain
the transmissivity distribution in FIG.6.

In FIG.7, we take $n_{b}(0)=n_{b}(b)=1.38$,
$n_{a}(0)=n_{a}(a)=2.35$, i.e., the transmissivity of conventional
photonic crystals. Compare FIG.6 with FIG.7, it can be found the
results: (1) when the line function of refractive indexes is up,
i.e., the GFPCs, the transmissivity can be far larger than $1$
($T$ maximum is about $850$), while the maximum of transmissivity
in conventional photonic crystals is $1$; (2) The number of band
gaps in GFPCs are more than the conventional PCs.

\begin{figure}[tbp]
\includegraphics[width=9 cm]{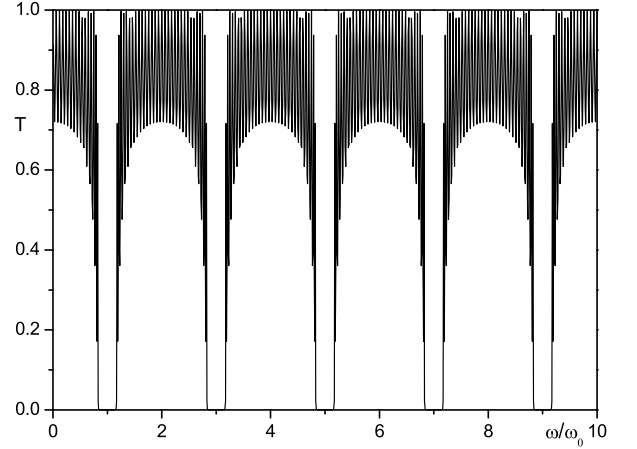}
\caption{The relation between transmissivity and frequency
corresponding to the conventional photonic crystals.}
\end{figure}

In FIGs.8-10, we main discuss the relation between transmissivity
and wave length corresponding to the GFPCs with defect layer,
there is $n_{d}d=\frac{\lambda_{0}}{4}$. Base on the structure of
GFPCs $(BA)^{16}$, we inset the defect layer at different
position, which are shown in FIG.8(a-c). The structures are:
(a)$(BA)^{8}D(BA)^{8}$, (b)$(BA)^{10}D(BA)^{6}$,
(c)$(BA)^{12}D(BA)^{4}$, where $n_{d}=2$. We can see that when the
position of defect layer is changed, the defect model become
small.
\begin{figure}[tbp]
\includegraphics[width=9 cm]{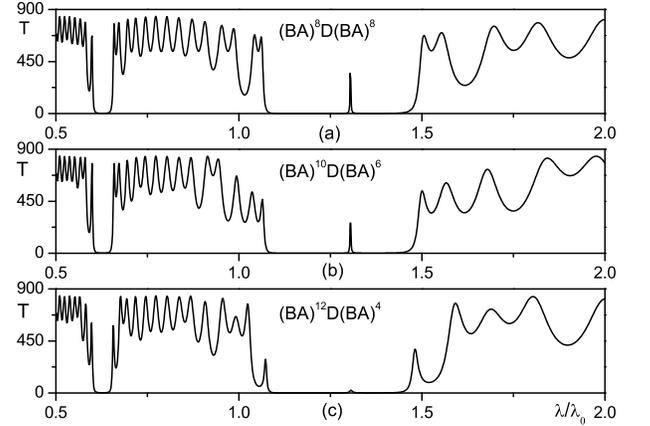}
\caption{Comparing the transmissivity of the GFPCs with different
position of defect layer.}
\end{figure}
\begin{figure}[tbp]
\includegraphics[width=8.5cm]{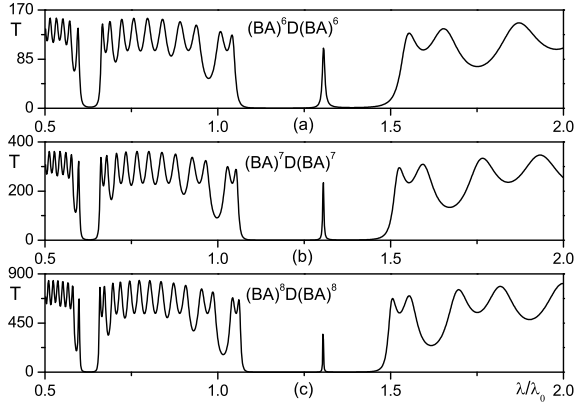}
\caption{Comparing the transmissivity of the GFPCs
($(BA)^{N}D(BA)^{N}$) with different half period numbers (a) N=6
(b) N=7 (c) N=8.}
\end{figure}
\begin{figure}[tbp]
\includegraphics[width=8.5 cm]{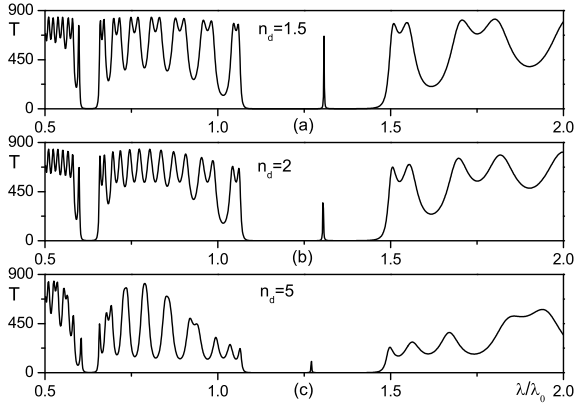}
\caption{Comparing the transmissivity of the GFPCs
($(BA)^{8}D(BA)^{8}$) with different refractive indexes of defect
layer.}
\end{figure}
In FIG.9, we compare the transmissivity with different structure
as $(BA)^{N}D(BA)^{N}$. In FIG.9(a-c), the half period number are:
$N=6$, $N=7$ and $N=8$, where $n_{d}=2$. We can obtain the
results: As the number of half period $N$ increase, for example,
$N=6$, the maximum of defect model achieve $110$ (FIG.9(a)). When
$N=7$, the intensity of defect model is about $250$ (FIG.9(b)).
When $N$ increases up to $8$ (FIG.9(c)), the maximum of defect
model nearly $400$, i.e., as the half period number $N$ increase,
the maximum of defect model increase, the defect model's width of
half height become narrow and the transmissivity of the GFPCs also
increase.

FIG.10 shows the transmissivity of the GFPCs ($(BA)^{8}D(BA)^{8}$)
with different refractive indexes of defect layer. From
FIG.10(a-c), $n_{d}$ are taken as: $1.5$, $2$ and $5$,
respectively. Here, we notice that as the refractive indexes of
defect layer increase, the intensity of defect model decrease, and
the position of defect model become blue shift.

\begin{figure}[tbp]
\includegraphics[width=8.5 cm]{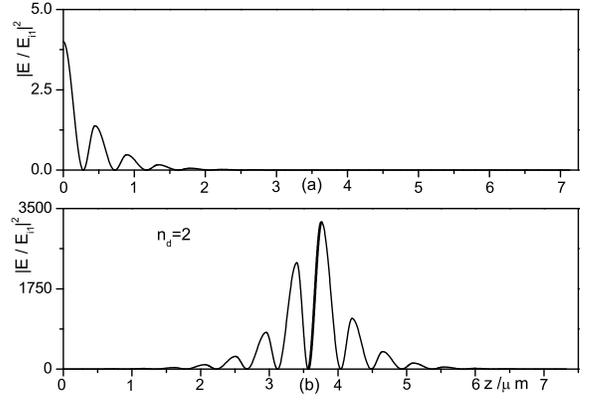}
\caption{The light distribution in the conventional PCs. (a)
without the defect layer($(BA)^{16}$), (b) with the defect layer
($(BA)^{8}D(BA)^{8}$). The bold line is the field distribution of
defect layer.}
\end{figure}
\begin{figure}[tbp]
\includegraphics[width=8.5 cm]{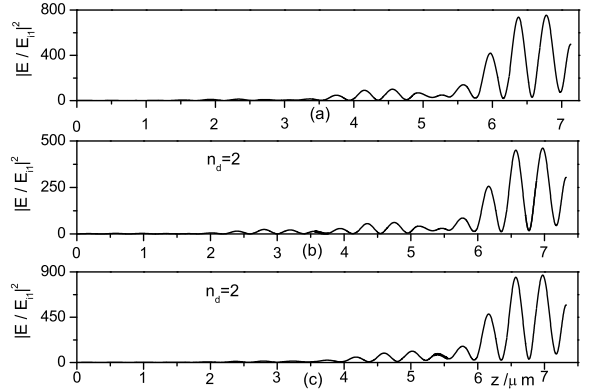}
\caption{The light distribution in the GFPCs. The structures are:
(a) $(BA)^{16}$, (b) $(BA)^{8}D(BA)^{8}$, (c)
$(BA)^{12}D(BA)^{4}$. The bold line is the field distribution of
defect layer.}
\end{figure}
\begin{figure}[tbp]
\includegraphics[width=8.5cm]{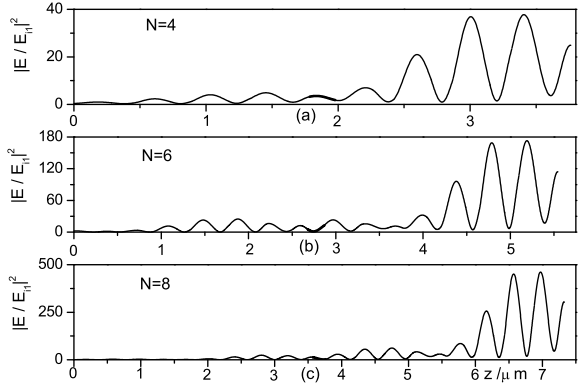}
\caption{The light field distribution of the GFPCs
($(BA)^{N}D(BA)^{N}$)with different periodicity, (a) N=4, (b) N=6,
(c) N=8. The bold line is the field distribution of defect layer.}
\end{figure}

FIGs.11-14 are the distribution of electric field. The transverse
axis $z$ is propagation distance, and the longitudinal axis is the
ratio of field intensity $E$ and incidence field intensity
$E_{i1}$ square, i.e., $|E/E_{i1}|^{2}$.

FIG.11 is the distribution of electric field in conventional PCs.
The structure of FIG.11(a-b) are $(BA)^{16}$ and
$(BA)^{8}D(BA)^{8}$. The main parameters are the same as FIG.7,
and the $n_{d}=2$, $n_{d}d=\frac{\lambda_{0}}{4}$. It can be found
that the electric field was enhanced obviously when inserted the
defect layer.

In the following, the main parameters are the same as FIG.6. In
FIG.12, we study the effect of different structure of GFPCs on the
distribution of light field. For FIG.12(a), the structure is
$(BA)^{16}$, FIG.12(b) and FIG.12(c), the structure are
$(BA)^{8}D(BA)^{8}$ and $(BA)^{12}D(BA)^{4}$, where $n_{d}=2$,
$n_{d}d=\frac{\lambda_{0}}{4}$. Comparing FIG.12(a) with
FIG.12(b), we found that when the defect layer is located in the
middle of GFPCs, the electric field was weaken. While in
FIG.12(c), the electric field was enhance. They are shown that
both the defect layer and its' position have the effects on the
distribution of light field, and we can find the defect layer made
the electric field local enhanced for the conventional PCs. While
the defect layer made the electric field whole enhanced or
decreased for the GFPCs.

In FIG.13, we shall consider the effect of half period number $N$
($(BA)^{N}D(BA)^{N}$) on the electric field. Taken $N=4$, $N=6$
and $N=8$ corresponding to FIG.13(a-c). The results shown that
when $N$ increase, the relative intensity of electric field
heighten.

FIG.14 show the effect of different refractive indexes of defect
layer ($n_{d}$) on the electric field. The structure is
$(BA)^{8}D(BA)^{8}$. In FIG.14(a-c), $n_{d}$ are equal to $1.5$,
$2$ and $3$, respectively. As we can see in FIG.14(a-c), when
$n_{d}$ increase, the relative intensity of electric field was
enhanced.

\vskip 2pt

{\bf 6. Conclusion}

\vskip 2pt
\begin{figure}[tbp]
\includegraphics[width=8.5 cm]{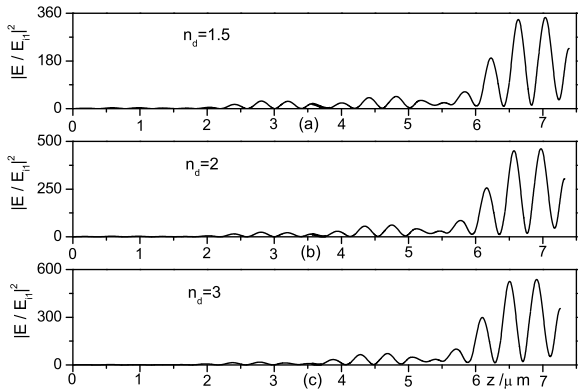}
\caption{The light field distribution of the GFPCs
($(BA)^{8}D(BA)^{8}$)with different refractive indexes of defect
layer. The bold line is the field distribution of defect layer.}
\end{figure}

In summary, We have theoretically investigated a new general
function photonic crystals (GFPCs) with defect layer. Based on
Fermat principle, we achieve the motion equations of light in
one-dimensional general function photonic crystals, and calculate
its transfer matrix. We choose the line refractive index function
for two mediums $A$ and $B$, and obtain some results: (1) When the
position of defect layer move behind in the GFPCs, or the
refractive indexes of defect layer increase, the transmission
intensity maximum of defect model decreases. (2) When the period
number of GFPCs increase, the transmission intensity of defect
model increase, and the defect model's width of half height become
narrow. (3) Both the defect layer and it's position have the
effects on the electric field. (4) In the structure
$(BA)^{N}D(BA)^{N}$, when the period number $N$ increase, the
relative intensity of electric field increase. (5) When the
refractive indexes of defect layer ($n_{d}$) increase, the
relative intensity of electric field was enhanced. Since the GFPCs
has new character different from the conventional PCs, it should
be helpful in the design of photonic crystals.

\end{document}